\documentclass[preprint,amssymb,prl,aps,superscriptaddress]{revtex4-1}
\usepackage{graphicx} \usepackage{amsmath}

\begin{document}


\title{Fresnel diffraction of spin waves}

\author{N. Loayza}
\affiliation {Colegio de Ciencias e Ingeniería, Universidad San Francisco de Quito, Quito, Ecuador}

\author{M. B. Jungfleisch}
\affiliation {Material Science Division, Argonne National Laboratory, Argonne, Illinois 60439, USA}
\affiliation {Department of Physics and Astronomy, University of Delaware, Newark, Delaware 19716, USA}

\author{A. Hoffmann}
\affiliation {Material Science Division, Argonne National Laboratory, Argonne, Illinois 60439, USA}

\author{M. Bailleul}
\affiliation {Institut de Physique et Chimie des Mat\'eriaux de Strasbourg, UMR 7504 CNRS, Universit\'e de Strasbourg, 23 rue du Loess, BP 43, 67034 Strasbourg Cedex 2, France}

\author{V.Vlaminck}
\affiliation {Colegio de Ciencias e Ingeniería, Universidad San Francisco de Quito, Quito, Ecuador}



\date{\today}

\begin{abstract}
The propagation of magnetostatic forward volume waves excited by a constricted coplanar waveguide is studied via inductive spectroscopy techniques. A series of devices consisting of pairs of sub-micrometer size antennae is used to perform a discrete mapping of the spin wave amplitude in the plane of a 30-nm thin YIG film. We found that the spin wave propagation remains well focused in a beam shape of width comparable to the constriction length and that the amplitude within the constriction displays oscillations, two features which are explained in terms of near-field Fresnel diffraction theory.  
\end{abstract}


\maketitle


The emerging field of magnonics \cite{Demokritov2012,Chumak2015} has sparked a renewed interest in non-uniform magnetization dynamics. Spin-waves are now considered as a very promising information carrier for performing basic logic operations \cite{Schneider2008,Chumak2014,Vogt2014,Klingler2015,Louis2016}, or implementing novel computation architectures \cite{Kozhevnikov2015, Papp2017}. A key advantage of spin-waves is that their dispersion can be easily tailored in a wide band of the microwave spectrum, particularly in the so-called magnetostatic wave regime for which the magnetic dipolar interaction plays the dominant role. Recently, it has been demonstrated that the propagation of spin waves in ferromagnetic thin film could be shaped using several concepts borrowed from optics \cite{Demidov2008}. A special attention has been set on understanding their refraction and reflection effects \cite{Gruszecki2014,Stigloher2016,Gruszecki2017,Gräfe2017}, and also on generating and manipulating spin-wave beams. The latter is of particular importance in order to exploit the potential of multi-beam interference. \\
So far, three different mechanisms have been investigated to shape spin-wave beams: (i) the so-called caustic effect \cite{Demidov2009,Schneider2010,Sebastian2013,JVKim2016} associated with the very strong anisotropy of magnetostatic wave dispersions; (ii) the confinement by the strongly inhomogeneous internal magnetic fields existing at strip edges \cite{Demidov2008}, in magnetic domain-walls \cite{Wagner2016}, or close to nano-contact spin torque nanoscillators \cite{Houshang2016,Demidov2016}; (iii) the coupling to specially designed constricted microwave antennae providing a suitable non-uniform magnetic pumping field profile \cite{Gruszecki2016,Körner2017}. The last method appears the most versatile, being able to produce a coherent spin wave beam in a homogeneous magnetic layer without any special requirement on its magnetic configuration. It was first proposed theoretically by Gruszecki et al. via micromagnetic simulation \cite{Gruszecki2016}, and was recently verified experimentally by K\"orner et al. via time resolved magneto-optical imaging of magnetostatic surface wave beams generated in a relatively thick NiFe film \cite{Körner2017}. In this letter, we show experimentally that the spin wave beam generated by a constricted coplanar waveguide (CPW) follows closely a near-field diffraction pattern. To this objective, we resort to all-electrical measurements performed in a configuration providing isotropic spin-wave propagation (thin Yttrium Iron Garnet film magnetized out-of-plane) and analyze them using elementary Fresnel diffraction modeling. \\


The spin wave antennae are designed in such a way that the constricted region of the CPW reproduces as closely as possible the case of an isolated rectangular slit. 
We first focus on the \textit{geometry-A} of spin-wave antennae shown in Fig.\,1-(a),(b). It consists of a pair of identical shorted CPWs, whose constriction is shaped symmetrically with a gradual bend in order to have two narrow sections of CPW facing each other. The constricted region of the CPW consists of a central track of width $w_S$\,=\,$400\,nm$ and two ground tracks of width $w_G$\,=\,$200\,nm$ with a gap of 200\,nm. The generated spinwaves have a wavelength of the order of the distance between the center of the ground tracks, i.e. $\lambda \simeq 1\,\mu m$ , which remains much smaller than the constriction length. We adopted a much sharper constriction than in the geometries study by Gruszecki et al. \cite{Gruszecki2016}, and K\"orner et al. \cite{Körner2017}, with a factor of ten between the widths of the constricted section I and the non-constricted section II. This allows us to fully separate the peaks associated with the excitation of spin waves in the two sections, as illustrated in Fig.\,1-(c) which shows the corresponding Fourier transforms of the current density (assumed to be uniform in each CPW track). For section I, one distinguishes a main peak centered at $k_{I}$\,=\,$5.92\,rad.\mu m^{-1}$ with a full width at half maximum $\Delta k_{I}$\,=\,$4.15\,rad.\mu m^{-1}$, and for section II a main peak at $k_{II}$\,=\,$0.59\,rad.\mu m^{-1}$ with $\Delta k_{II}$\,=\,$0.41\,rad.\mu m^{-1}$. \\
We fabricated five spin-wave transducers of $geometry-A$ with separation distance $D$\,=\,$\{4,6,8,10,12\}\,\mu m$ and constriction length $l_{exc}$\,=\,$5\,\mu m$, which we used for preliminary characterization and validation of the spin-wave transduction in continuous layer. The antennae were fabricated by e-beam lithography and lift-off of $5\,nm\,Ti\,/\,80\,nm\,Au$ directly on top of $30\,nm$ thin sputtered YIG (Y$_{3}$\,Fe$_{5}$\,O$_{12}$) films deposited on gadolinium gallium garnet by magnetron sputtering and post-annealed \cite{Joungfleisch2017,Li2016,SOM_S1}. The fabricated device is then placed in the center of the lower pole of an electromagnet fitting in a home-made probe station, and we proceed to the propagative spin-wave spectroscopy measurement \cite{Vlaminck2010} while applying an external magnetic field $H$ large enough to magnetize the film out of the plane. This corresponds to the so-called magnetostatic forward volume wave (MSFVW) configuration, for which the isotropic dispersion relation does not favor any propagation direction. This is in strong contrast with the situation of in-plane magnetized films, for which the spin-wave dispersion is strongly anisotropic with a maximal group velocity in the so-called magnetostatic surface wave configuration. For practical reasons, most studies of nanomagnonics, including recent ones in YIG have been done in this last configuration \cite{Maendl2017,Collet2017}. In the present case, we simplify the analogy with optics by employing the isotropic MSFVW configuration, and take directly advantage of the low damping and low magnetization of the YIG films.

\begin{figure}
\includegraphics[width=\linewidth]{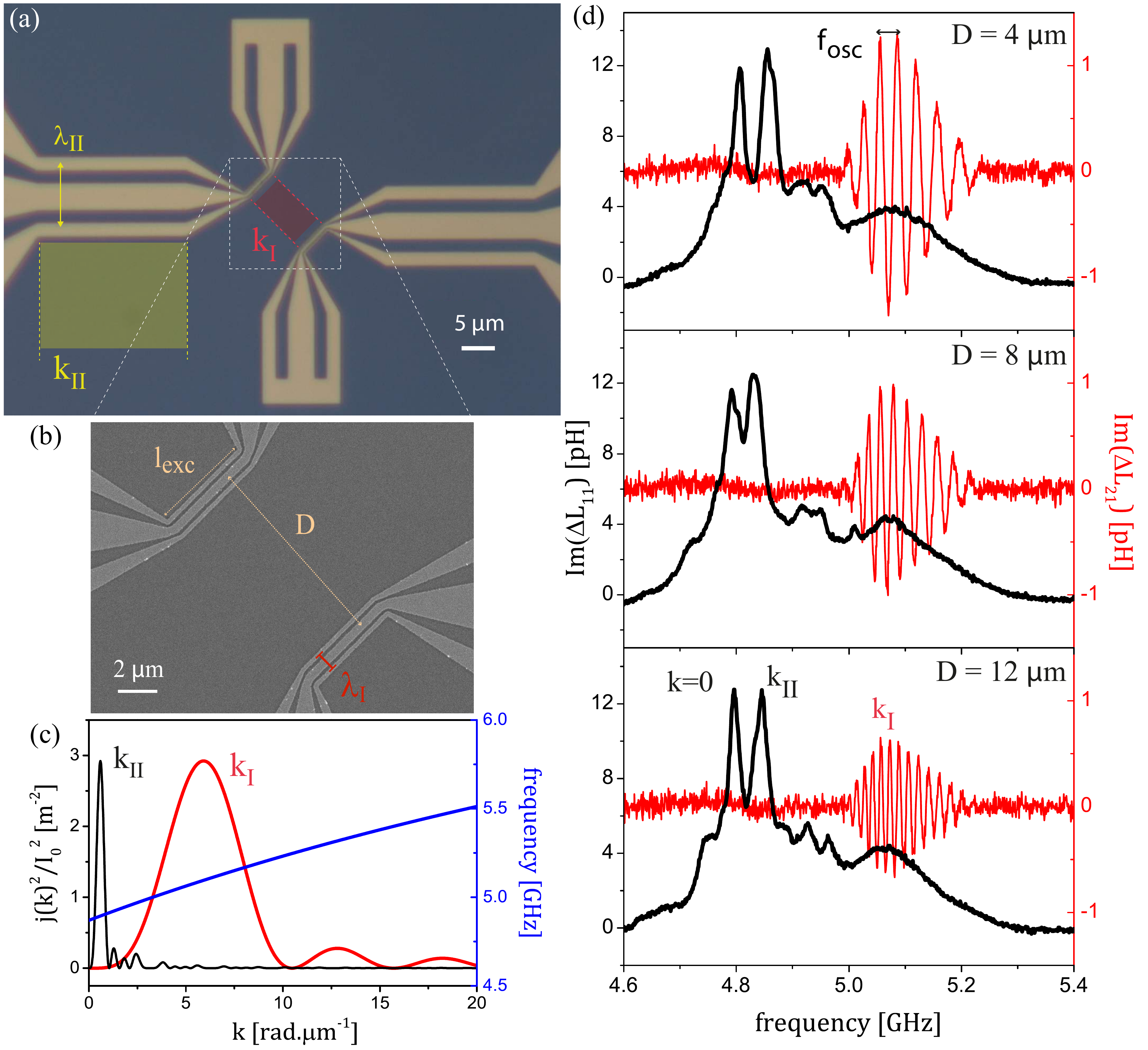}
\caption{(a) \textit{Geometry-A} spin-wave antennae with a separation distance $D$\,=\,$12\,\mu m$. (b) Scanning electron microscope image of the same device zoomed in the region of the constriction. (c) Fourier transform of the current distribution for the two sections I and II, and MSFVW dispersion relation for $\mu_0\,H_{ext}$\,=\,$308\,mT$. (c) Self- and mutual inductance spectra obtained at $\mu_0\,H_{ext}$\,=\,$308\,mT$ for three devices of \textit{geometry-A} with separation distance $D$\,=\,$4\mu m$, $D$\,=\,$8\mu m$, and $D$\,=\,$12\mu m$.}
\label{Fig1}
\end{figure}

The microwave spectra were acquired using a vector network analyzer ($Agilent\,E8342B$, $10\,MHz$\,-\,$50\,GHz$) at low input power ($-20\,dBm$), $100\,Hz$ bandwidth, and in a single sweep mode in order to limit the possible temperature drift of the electromagnet. We always perform two measurements: a first one at a resonant field ($H_{res}$), followed by a second one at a reference field ($H_{ref}$) for which no resonance occurs within the frequency range swept. In this manner, we retrieve the variation of inductance $\Delta\,L_{ij}$\,=\,$L_{ij}(H_{res})-L_{ij}(H_{ref})$ due to spin wave excitation \cite{Collet2017}; $\Delta\,L_{11}$ the self-inductance measured on antenna $1$, and $\Delta\,L_{21}$ the mutual inductance characterizing the transduction of spin wave excited by antennae 1 and detected by antennae 2. Figure\,1-(d) shows typical spectra obtained with identical antennae of \textit{geometry-A} at $H_{ext}$\,=\,$308\,mT$, for three different separation distances ($4$, $8$, and $12\,\mu\,m$). We can identify from the reflection spectra three main peaks which are attributed to the different parts of the CPW. Namely, the lowest frequency peak corresponds to the quasi-uniform resonance ($k$\,=\,$0$) of the wide section of the CPW where the $150\,\mu\,m$ pitch coplanar probe are contacted. The second peak corresponds to the non-constricted region II of the CPW, and the last peak to the constricted region I. For the mutual inductance spectra, we observe oscillations only underneath the last peak confirming that only the constricted region of the CPW contributes to the spin wave transduction between antennae. These oscillations are attributed to the phase delay $kD$ accumulated by the spin-waves during its propagation between the two antennas,  
and therefore are more numerous the longer the separation distance between antennae. The level of amplitude of the mutual inductance spectra is comparable to the one found when performing simulation of MSFVW transduction \cite{Vlaminck2010,SOM_S2} on a stripe of width equal to the length of the constricted region. This suggests already that the excitation of the spin wave from this type of constriction should remain fairly focused. \\ 
\begin{figure*}
\includegraphics[width=\linewidth]{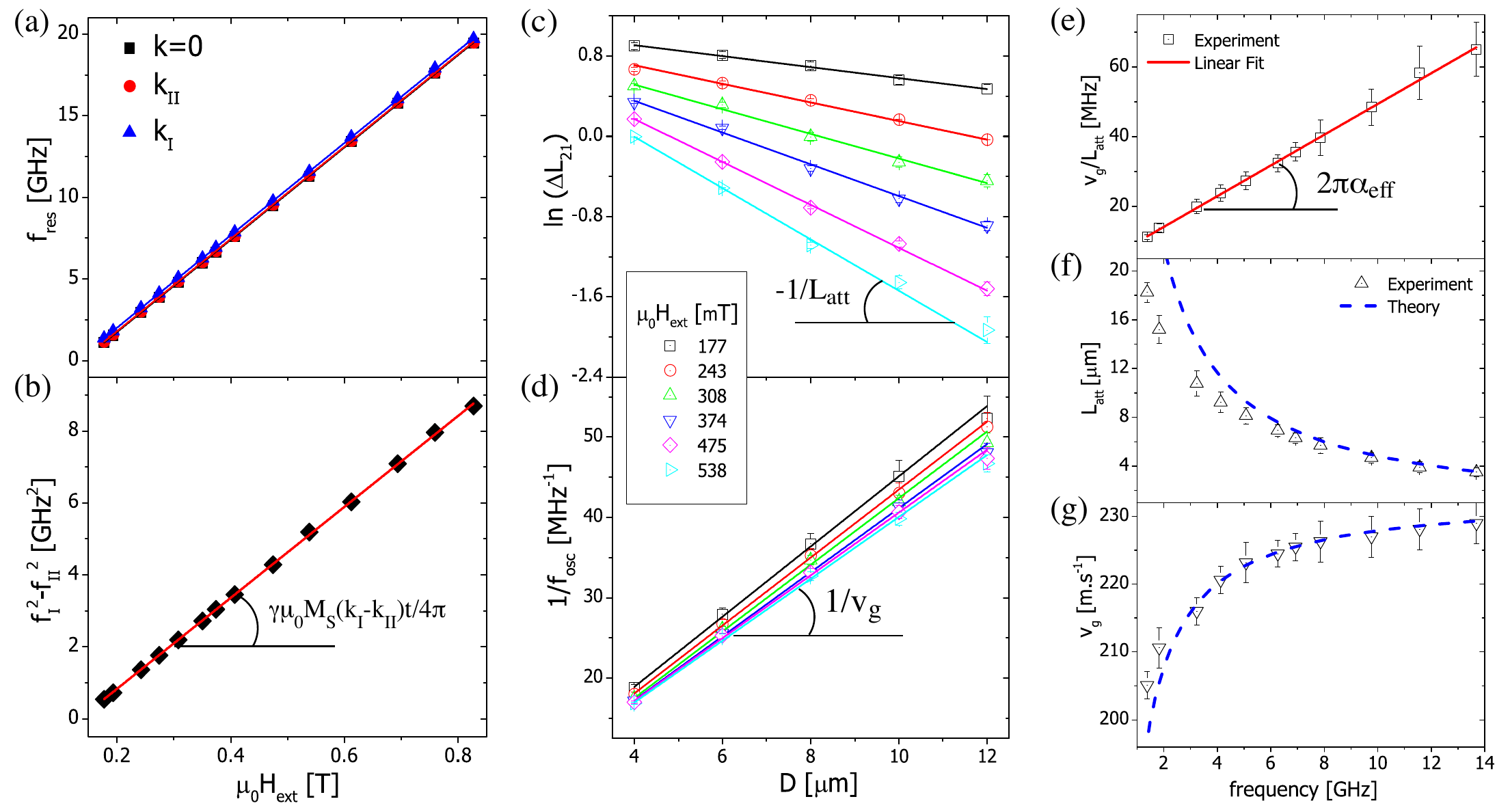}
\caption{(a) Field dependence of the resonance peaks $k$\,=\,$0$, $k_{II}$, and $k_I$. (b) Difference of the square of the resonance frequencies $f^2(k_I)$\,-\,$ f^2(k_{II})$. Separation distance dependence of (c) $ln(\Delta\,L_{21})$, and (d) the inverse of the oscillation period of $\Delta\,L_{21}$. Frequency dependence of: (e) the ratio of the group velocity to the attenuation length, (f) the attenuation length, and (g) the group velocity. }
\label{Fig2}
\end{figure*}
To validate our spin wave transduction technique when applied to a continuous magnetic layer, we first analyze the microwave spectra measured for different fields and different distances between the antennas. In particular, we can take advantage of the three section of our waveguides to perform k-resolved ferromagnetic resonance (FMR). As shown in Fig.\,2-(a), we track the peak position in function of applied field respectively for $k$\,=\,$0$, $k_{II}$, and $k_{I}$. Fitting it to the MSFVW dispersion relation \cite{Gurevich1996} for five different devices, we obtain an average value for the gyromagnetic ration $\gamma/2\pi$\,=\,$28.26\,\pm\,0.07\,GHz.T^{-1}$ and the effective magnetization $\mu_{0} M_{eff}$\,=\,$136 \pm 2\,mT$. Furthermore, we can estimate independently the saturation magnetization $M_{s}$ by plotting the field-dependence of the difference $f_{res}^{2}(k_{I})$-$f_{res}^{2}(k_{II})$\,=\,$\gamma^{2} \mu_{0}^{2} (H_{ext}$\,-\,$M_{eff}) M_{s} (k_{I}-k_{II}) t/2$ (where $t$ is the YIG film thickness) as shown in Fig.\,2-(b), from which we find a nice linear dependence and the average value $\mu_{0} M_{s}$\,=\,$196 \pm 8\,mT$. Next, we use the observed decay of the amplitude of the mutual-inductance as a function of the distance as shown in Fig.\,2-(c) to extract the characteristic attenuation length of the spin wave. 
For each applied field, we observe a clear linear dependence of $ln(|\Delta\,L_{21}|)$ on the antenna separation $D$, which is consistent with an exponential decay  $|L_{21}| \propto e^{-D/L_{att}}$). This constitutes another evidence for a proper focusing of the spin wave excitation. Obviously, a diffused emission of opening angle $\theta$ would reduce the amplitude by an additional factor $l_{det}/(D\,\theta)$, which is not observed here. Then, from the period of oscillation $f_{osc}$ of the mutual inductance spectra, we can estimate the group velocity $v_g$ according to $v_g$\,=$\,f_{osc}\,D$ \cite{Vlaminck2010}. Fig.\,2-(d) shows clear linear dependence of $1/f_{osc}$ with $D$ for the different applied fields. Finally, we perform a linear fit of the frequency dependence of the ratio $v_g/L_{att}$ \cite{Gladii2016},and identify the slope to $2 \pi \alpha_{eff}$ [cf. Fig.\,2-(e)], which gives us a value of the effective damping $\alpha_{eff}$\,=\,$7.5 \pm 0.2\,10^{-4}$ in good agreement with previous measurements on similar films \cite{Liu2014,Joungfleisch2015}. 
We obtain fairly good agreements with the theoretical group velocity and attenuation length estimated from the MSFVW dispersion relation [dotted lines in Fig. 2(f,g)], which validate the implementation of the spin wave transduction technique to continuous layers for this geometry of CPW.


We now turn to the main result of this work, which is the evolution of the amplitude $\Delta\,L_{21}$ between several pairs of antennae at various separation distances $D$, and with various shift $s$ with respect to their axis in order to map in a discrete manner the spin-wave emission from a constriction. We fabricated two series of pairs of non-identical spin-wave antennae with a long excitation antenna ($l_{exc}$\,=\,$10\,\mu m$), and a shorter detection antenna of ($l_{det}$\,=\,$2\,\mu m$) in order to refine the spatial resolution of the mapping. 
The first series consists of the symmetrical \textit{geometry-A} as shown in Fig. 3-(a), for which we fabricated six devices having the same separation distance $D$\,=\,$5\,\mu m$, and only one-sided shift $s$\,=\,$\{0,1,2,3,4,6\}\,\mu m$. 
For the second series, \textit{geometry-B} shown in Fig.\,3-(b), which consists of an asymmetrical constriction short-circuited right at its end and also with a steeper bend, we fabricated eighteen devices covering two separation distances $D$\,=\,$\{8,12\}\,\mu m$, and nine shift $s$\,=\,$\{-8,-6,-4,-2,0,2,4,6,8\}\,\mu m$. 
Fig.\,3-(c) shows the shift dependence of the peak amplitude $|\Delta\,L_{21}|_{max}$ for \textit{geometry-A} at various applied field  (see typical examples of mutual-inductance spectra in the supplementary materials \cite{SOM_S3}). We observe an oscillation of the amplitude within the width of the constriction and a clear drop of amplitude for the device $s$\,=\,$6\,\mu\,m$, which lays just entirely outside of the constriction. Similar observations are made with \textit{geometry-B} shown in Fig.\,3-(d) although the drop of amplitude outside the constriction is slower for negative shifts due to the non-symmetrical shape of the antennae. Indeed, the shorted ends of the constrictions, which come close to each other for positive $s$ [see Fig. 3(b)], radiate much less spin-wave power out of the constriction than the broader convex CPW access, which come close to each other for negative $s$.\\

\begin{figure}
\includegraphics[width= \linewidth]{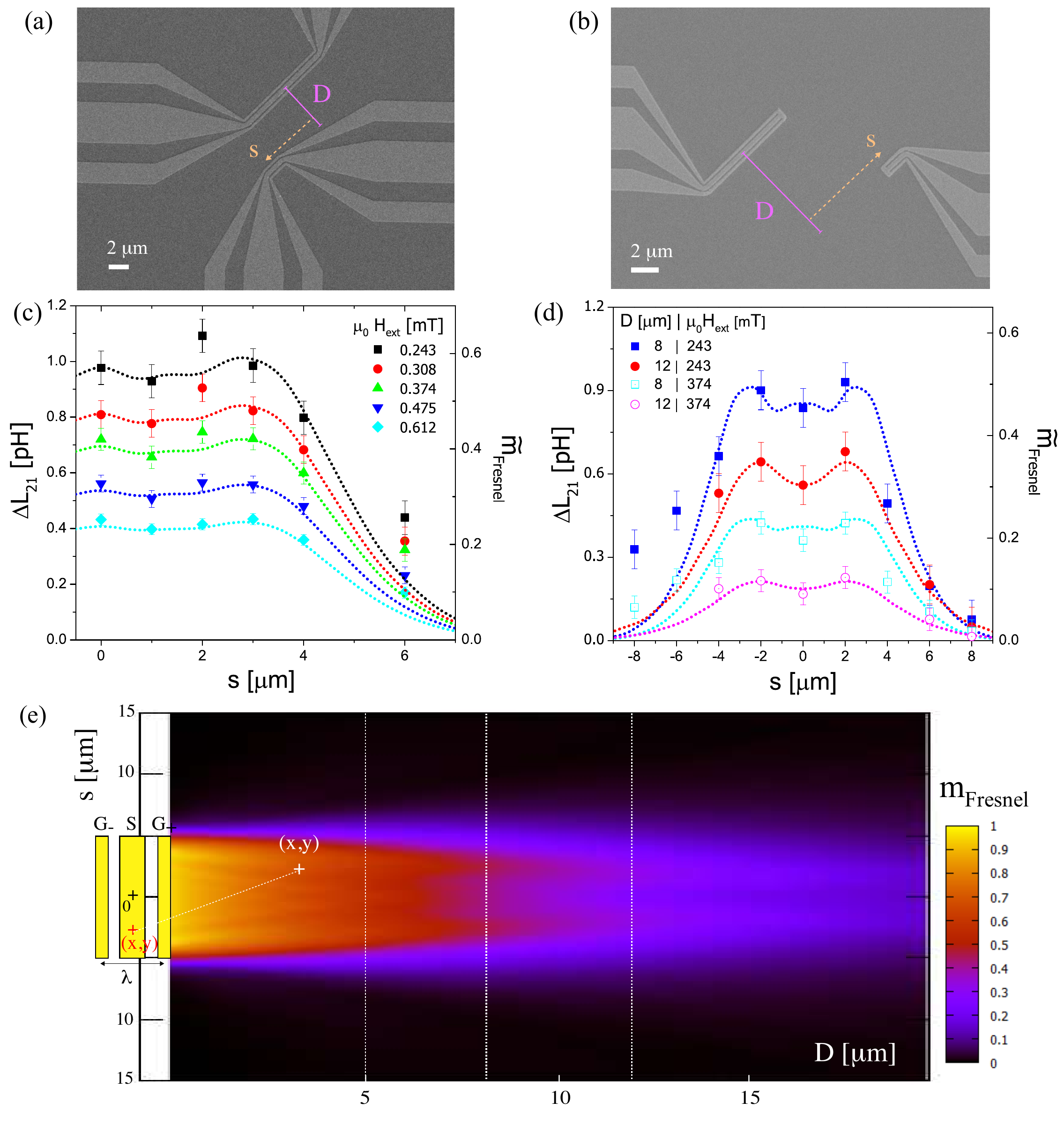}
\caption{(a) \textit{Geometry-A} device for the mapping of spin wave with a separation distance $D$\,=\,$5\mu m$ and a shift $s$\,=\,$-5\mu m$. (b) \textit{geometry-B} with a separation distance $D$\,=\,$12\mu\,m$ and a shift $s$\,=\,$+8\mu\,m$. Evolution of the measured mutual inductance amplitude with the antennae shift $s$ for: (c) \textit{geometry-A} mapping devices, and (d) \textit{geometry-B} antennae separated by $D$\,=\,$8\mu\,m$, and $D$\,=\,$12\mu\,m$. The dotted lines are the calculated spin-wave amplitude from the Fresnel diffraction model with the corresponding $L_{att}$. The symbols are the measured ampitude for the different devices. (e) Color mapping of the wave amplitude for $L_{att}$\,=\,$10\,\mu\,m$ taking into account the probe size $l_{det}$\,=\,$2\,\mu\,m$. The vertical dotted lines indicate sections of amplitude $<\tilde{m}_{CPW}(s,D)>$ at $D$\,=\,${5,8,12}\,\mu\,m$,}
\label{Fig3}
\end{figure}
To describe these features of spin wave emission from a constricted CPW, we propose to implement the common equations of optics used in the case of the Fresnel diffraction from a rectangular slit \cite{Hecht2002}. This choice is particularly relevant for the range of wavelength considered, for which the Fresnel radius $R_F$ remains much smaller than the length of the constriction: $R_F$\,=\,$\sqrt{\lambda\,D}<<l_{exc}$. We simplify the problem by considering that each track of the CPW [$j$\,=\,$\{G_-,S,G_+\}$, see sketch in Fig.\,3-(e)] acts a single rectangular source of coherent, circular, and monochromatic waves, of wavelength $\lambda$\,=\,$\dfrac{2\pi}{k_I}$. We also account for the spin wave attenuation with an exponential decay factor ($e^{-r/L_{att}}$), where $L_{att}$ is field- (or frequency-) dependent with a value given in Fig.\,2-(f). The normalized spin-wave amplitude $\tilde{m}_{Fresnel}(D,s)$ at a distance $D$ and a shift $s$ emitted by the track $j$ of the CPW is written as:
\begin{align}
\textstyle \tilde{m}_{j} (D,s) = \int_{-l/2}^{l/2} dy \int_{-w_j/2}^{w_j/2} dx \dfrac{1}{\sqrt{r_j}} e^{-r_j/L_{att}} e^{-ikr_j}
\end{align}
Where  $r_j$\,=\,$\sqrt{(D-x_j)^2+(s-y)^2} $ is the distance between an element of surface $dxdy$ of the source centered at $(x_j,y)$ and a detection point of coordinates $(D,s)$; $l$ is the antennae length and $w_j$ the width of the CPW track. Now, the complete amplitude of the Fresnel diffracted spin-wave $\tilde{m}_{CPW}$ results from the linear combination of the three branches of the CPW: 
{\small
\begin{align}
 \textstyle \tilde{m}_{CPW} (D,s) = -\tilde{m}_{G_-}(D+\dfrac{\lambda}{2},s) + \tilde{m}_{S}(D,s) - \tilde{m}_{G_+} (D-\dfrac{\lambda}{2},s)      
\end{align}}
Where the negative signs accounts for the opposite phase of the excitation in the ground lines with respect to the central line. 
Fig.\,3-(e) shows the color mapping of the spin-wave amplitude in the $(D,s)$ plane calculated from Eq.\,(2) with an attenuation length $L_{att}$\,=10\,$\mu$\,m. In order to compare our measurement with this Fresnel diffraction model, we took into account the non-punctual aspect of the detection antenna by averaging the amplitude over the probe antenna extension ($l_{det}$\,=2\,$\mu$\,m). This near-field diffraction patterns reproduces the main features of our measurement, which are on one hand an emission that remains focused in a beam shape of width similar to the CPW length, and on the other hand, some oscillations of the amplitude within the beam width that depend mostly on the distance D. 
Finally, we compare the measured amplitudes with the calculated ones $<$\,$\tilde{m}_{CPW}(s,D)$\,$>$ [dotted lines in Fig.\,3-(c),(d)] for the specific distances $D$, and with the corresponding values of attenuation length obtained in Fig.\,2-(f). We find a remarkable agreement between this Fresnel diffraction model of spin waves emitted from an antenna of finite extension and our measurements in the two different geometries of waveguide, which constitutes a direct demonstration of the focused nature of spin wave beams in constricted CPW. \\

In summary, we first demonstrated the possibility of performing spin-wave spectroscopy in thin magnetic films without the need to structure a spin-wave guide, only by using sufficiently sharp constrictions in CPWs. 
We firstly showed that the signal amplitudes measured for pairs of identical antennae shifted gradually along the beam direction follow precisely an exponential decay, which suggests that the emission remains well-focused. Secondly, via a series of devices consisting of pairs of non-identical antennae covering different location of the 2D-plane, we performed a discrete mapping of the spin-wave amplitude for two different geometries conceived in such a way to reproduce the case of an optical rectangular slit. We found that the spin wave amplitude oscillates within the constriction zone, while it decays rapidly outside of it, which is notably well-explained with a Fresnel diffraction model of circular waves. These findings draw a deeper parallel between the excitation of spin-waves from sub-micrometric antennae and the basic concepts of optics, and therefore pave the way for future studies of spin wave beam interference, which could find applications for spin wave logic devices.
\\
\\
We thank Olga Gladii, Hicham Majjad, Romain Bernard, and Alain Carvalho for support with the nanofabrication  in the STnano platform, and Guy Schmerber for X-ray measurements. This work was supported by the French “Agence National de la Recherche” grant ANR-11-LABX-0058\_\,NIE, and the USFQ´s $PolyGrant\sharp431$ program. The synthesis of the YIG films at Argonne was supported by the U.S. Department of Energy, Office of Science, Materials Science and Engineering Division.
\\

\end{document}